\documentclass[sigconf]{acmart}

\setcopyright{none}
\acmConference{}{}{}
\acmBooktitle{}
\acmPrice{}
\acmDOI{}
\acmISBN{}
\renewcommand\footnotetextcopyrightpermission[1]{}

\usepackage{listings}
\usepackage{enumitem}
\usepackage{graphicx}
\usepackage{float}
\begin{document}

\title{Securing High-Performance Data Transfers: Implementing AES Encryption in RDMA Systems}

\author{Erik Bångsbo}
\authornote{Both authors contributed equally to this research.}
\email{bangsbo@chalmers.se}
\author{Zakaria Hersi}
\authornotemark[1]
\email{hersiz@chalmers.se}
\affiliation{%
  \institution{Chalmers student and Saab intern}
  \city{Gothenburg}
  \country{Sweden}
}

\author{Anna Benktson}
\email{anna.benktson2@saabgroup.com}
\author{Stefan Holmgren}
\email{stefan.holmgren@saabgroup.com}
\affiliation{%
  \institution{Saab AB}
  \city{Gothenburg}
  \country{Sweden}
}

\author{Romaric Duvignau}
\email{duvignau@chalmers.se}
\affiliation{%
  \institution{Chalmers University of Technology and University of Gothenburg}
  \city{Gothenburg}
  \country{Sweden}
\email{duvignau@chalmers.se}
}

\begin{abstract}
Remote Direct Memory Access (RDMA) is a key enabler of high-performance systems, offering low latency, high throughput, and reduced CPU overhead by allowing direct memory-to-memory transfers between machines. However, its design bypasses traditional CPU-mediated security mechanisms, introducing critical vulnerabilities in untrusted environments.
This work explores the integration of RDMA and AES-128 encryption to secure data transfers without compromising performance. We implement encryption directly within the data plane of a programmable Tofino switch using the P4 programming language. By offloading encryption from the CPU to the switch, our design preserves RDMA’s performance benefits while addressing its security shortcomings.
Experimental results show that the system achieves throughput of 0.37~Gbps for 16-byte packets, 0.76~Gbps for 32-byte packets, 1.83~Gbps for 64-byte packets, and 1.9~Gbps for 128-byte packets. These findings demonstrate the feasibility of secure, high-throughput RDMA communication using programmable network hardware.
\end{abstract}


\maketitle

\lstset{
  language=C,
  basicstyle=\ttfamily\scriptsize,
  keywordstyle=\color{blue},
  commentstyle=\color{gray},
  stringstyle=\color{red},
  morekeywords={header, struct},
  escapeinside={(*@}{@*)}, 
  numbers=left,
  numberstyle=\tiny\color{gray},
  frame=single,
  rulecolor=\color{black},
}

\section{Introduction}
Remote Direct Memory Access (RDMA)~\cite{kalia2014using, guo2016rdma, kalia2016design},  is a key technology in modern data centers and HPC environments. By enabling direct memory-to-memory transfers without CPU intervention, RDMA provides low-latency, high-throughput communication~\cite{planeta2023cord} for applications such as distributed storage, machine learning, and real-time analytics. This makes RDMA integral to performance-critical infrastructure.
Despite its performance advantages, RDMA lacks built-in support for encryption or authentication. While beneficial for performance, this design introduces significant risks in multi-tenant or untrusted environments. Attackers can exploit RDMA’s direct memory access to read or tamper with data in transit. As a result, RDMA adoption remains limited in domains requiring strong security.

Traditional RDMA security approaches rely on host-side encryption via CPUs or NICs (see e.g.~\cite{SecuringRDMA}). However, they either reduce RDMA’s performance benefits or require higher-end custom hardware. On the other side, programmable-switch dataplane encryption has been demonstrated on Tofino ~\cite{chen2020implementing}, but only for fixed-size payloads and not for RDMA traffic.
Hence, this paper addresses the research question: \textbf{can we secure RDMA traffic using programmable switches—without sacrificing throughput or increasing CPU load?} We answer this by presenting and evaluating a practical PoC system that integrates AES-128 encryption directly into the data plane of a Tofino switch using the P4 language.

\vspace{0.5em}
\noindent\textbf{State of the art.} 
In programmable switches, even basic functions such as forwarding must be implemented in code. 
Establishing RDMA connectivity therefore requires implementing RDMA in P4. Previous work has shown how to do it for RoCEv1~\cite{beltman2020using} and RoCEv2~\cite{xing2022bedrock}. However, none of these have also implemented encryption. This becomes a problem of integration, where the correct handling of RDMA needs to work in a single program with a solution that can encrypt traffic. 
%
Prior efforts to secure high-performance data transfers have primarily focused on NIC-based encryption for RDMA, or on implementing AES within programmable switches. 
For example, on one hand, Taranov et al.~\cite{ReDMArkBypassing} developed a NIC-based encryption mechanism that achieves low latency and high throughput for RDMA traffic. However, their approach is tightly coupled to specific hardware and incompatible with standard protocols like IPsec~\cite{hauser2020p4}, whose performance overhead makes it unsuitable for RDMA’s stringent speed requirements.
On the other hand, Chen~\cite{chen2020implementing} implemented AES encryption directly in the Intel Tofino switch data plane, tailoring it to the constraints of programmable switch architectures. While this solution effectively encrypts fixed 16-byte payloads, it is limited in flexibility and does not generalize well to real-world RDMA use cases that require variable packet sizes and more adaptable handling.



\vspace{0.5em}
\noindent\textbf{Contributions.} This work is thus the first to extend AES-based dataplane encryption specifically to high-throughput/low-latency RDMA traffic, making the following contributions:
\begin{itemize}[leftmargin=5.5mm]
    \item We design and implement a dataplane encryption pipeline that supports AES-128 encryption for RDMA traffic, handling variable payload sizes in the Tofino architecture using P4.
    \item We overcome key engineering challenges: strict switch resources and pipeline constraints, and full RDMA support.
    \item We conduct a thorough performance evaluation, demonstrating that our solution achieves up to 1.9~Gbps throughput with minimal packet loss ($<$0.001\%) even at smaller packet sizes, making it viable for production use.
\end{itemize}

While high-end RDMA NICs offer native encryption, they are costly and introduce vendor lock-in. In contrast switch based encryption offers a compelling alternative by enabling device agnostic protection for packets from multiple sources. By centralizing the encryption process within a single unit, it simplifies system design and reduces the reliance on specialized hardware. Our work presents a cost-effective, centralized security PoC where the switch acts as a gatekeeper for hosts designated for other purposes. Furthermore, using pure P4 ensures architectural portability; the logic can be compiled to any P4-target.
While Intel has been phasing out Tofino products from 2025 onwards, we believe that the mechanisms developed in our P4 implementation remain broadly applicable. In particular, they offer value for future switch-based processing and for the design of novel networking solutions, such as eBPF programs~\cite{hoiland2018express} running within the Linux kernel.
Our results indicate that secure, high-throughput RDMA transfers are achievable in the dataplane without involving the host CPU, thereby preserving the original performance benefits of RDMA. This has important implications for future secure computing infrastructure in cloud and edge data centers.

\vspace{0.5em}
\noindent\textbf{Threat model.}
We assume an adversary who can observe, replay, inject, and modify packets on the network path between RDMA hosts and the programmable switch. The primary assets are the confidentiality and integrity of RDMA traffic and the secrecy of cryptographic keys. The proposed design mitigates passive eavesdropping by offloading AES-128 encryption to the data plane, however, active attacks (e.g. replay or packet tempering) 
are not considered here.

\vspace{0.5em}
\noindent\textbf{Paper outline.}
Section~\ref{sec:background} reviews RDMA and P4 switches. Section~\ref{sec:design} presents the design and implementation of our dataplane encryption system. Section~\ref{sec:evaluation} evaluates performance. Finally, Section~\ref{sec:conclusion} concludes insights and directions for future improvement.

\section{Background} \label{sec:background}

We provide here a brief overview of RDMA architecture 
alongside the architecture of programmable switches, relevant network programming tools and security mechanisms.

\begin{figure}[t]
  \setlength{\abovecaptionskip}{2pt}   
  \setlength{\belowcaptionskip}{0pt}   
  \centering
  \includegraphics[trim={700 450 20 500},clip,
                   page=1,width=0.7\linewidth]{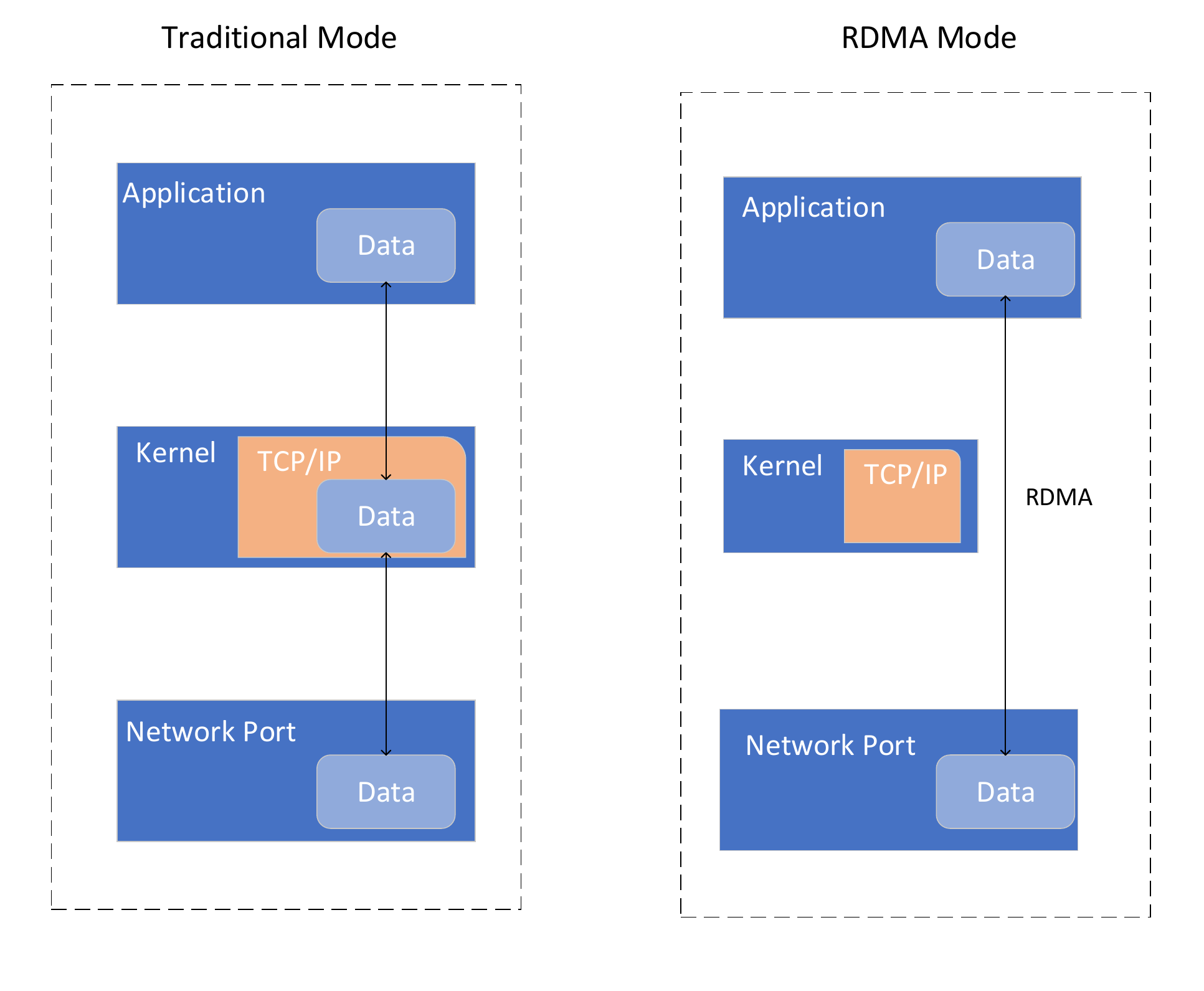}
  \caption{Diagram illustrating the differences between Traditional Mode vs.\ RDMA Mode.}
  \label{fig:RDMAdiagram}
\end{figure}

\subsection{Remote Direct Memory Access}

Remote Direct Memory Access (RDMA) is a key technology in HPC that enables direct memory access between servers NICs and memory, bypassing the CPU (see Figure~\ref{fig:RDMAdiagram}'s illustration). This architecture delivers $\mu$s-level latency and data transfer speeds exceeding 100 Gbps, significantly outperforming traditional TCP/IP-based networking \cite{planeta2023cord}. 
RDMA lacks built-in encryption, necessitating additional security measures for broader deployment~\cite{SecuringRDMA}.
\textit{RoCEv2} (RDMA over Converged Ethernet v2) adapts RDMA for data centers by running over UDP/IP, providing scalability and compatibility with standard Ethernet infrastructure \cite{RoCEV2_Survey}.
Among its transport modes, \textit{Reliable Connection} (RC) is connection-oriented, guaranteeing in-order, reliable delivery between paired queue pairs (QPs) and supports Send/Receive, Write, Read, and Atomic operations while \textit{Unreliable Datagram} (UD) is connectionless, with no guarantees on delivery or order and only supports Send/Receive operations.
The RoCEv2 packet adds a Basic Transport Header (BTH), an Invariant Cyclic Redundancy Check (ICRC), and a Frame Check Sequence (FCS) to the standard UDP packet structure. 

\subsection{Intel Tofino Switch Architecture and P4}

\begin{figure}[t]
  \setlength{\abovecaptionskip}{2pt}
  \setlength{\belowcaptionskip}{0pt}
  \centering
  \includegraphics[trim={40 40 40 60cm},clip,
                   page=2,width=0.82\linewidth]{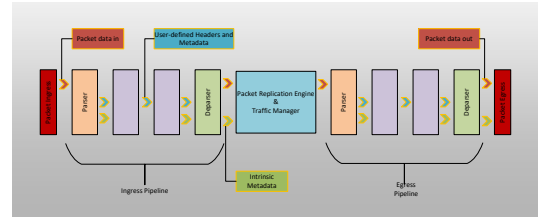}
  \caption{The pipeline architecture for TNA.}
  \label{fig:ArchTNA}
\end{figure}

Intel’s Tofino ASIC is a programmable Ethernet switch built on the Protocol Independent Switch Architecture (PISA), programmable using the P4 language \cite{agrawal2020intel}. 
%
The Tofino architecture integrates multiple components—e.g. the Intelligent Fabric Processor, FPGAs/IPUs, and Xeon processors—to offer a versatile networking platform. 
%
Although Tofino lacks dedicated cryptographic hardware, it is known to offer the possibility of AES encryption directly within the \textit{data plane} \cite{chen2020implementing}, i.e., the logical plane responsible of handling low-level protocols such as Ethernet, IP, UDP/TCP, and responsible for packet forwarding. 
Tofino Native Architecture (TNA) offers a P4 programming interface closely aligned with Tofino’s hardware features, enabling efficient and customized packet processing (see Figure~\ref{fig:ArchTNA}).

P4 (\url{https://p4.org/}) is a domain-specific language designed to program the data plane of network devices. 
It provides protocol-independent packet processing, allowing engineers to define headers and processing logic without being bound to specific protocols. 
Standard or custom packet headers can be defined within the language (via the \texttt{header} keyword specifying named data structures composed of typed fields with fixed bit widths, e.g., \texttt{bit<48>} for Ethernet addresses).
First, a parser extracts headers from incoming packets via a state machine.
After parsing, packet processing is driven by match-action tables that match packet header fields against entries and execute corresponding actions (e.g. modify headers, forward packets, or drop them).
A deparser concludes the pipeline where packets are emitted through output ports.
See~\cite{hauser2023survey,tan2021in,Kfoury2021ExhaustiveSurveyP4} for a more comprehensive presentation of P4 and P4's security.

\subsection{Method of Encryption}
This work focuses on implementing AES-128~\cite{dworkin2001advanced}, a widely used symmetric encryption standard in secure communications.
We explore practical AES encryption within programmable network devices, using Scrambled Lookup Tables to overcome the absence of dedicated cryptographic co-processors~\cite{chen2020implementing}.
AES-128’s balance of hardware efficiency and adaptability makes it a standard choice for FPGA and ASIC implementations (cf.~\cite{CryptoChap10}). 
%
AES is a block cipher with several modes of operation offering different security and performance trade-offs. 
We implement here the \textit{Electronic Codebook} (ECB) which encrypts each block independently by XORing plaintext with the key. Identical plaintext blocks yield identical ciphertexts, making ECB vulnerable to pattern-based attacks and is generally not recommended~\cite{blazhevski2013modes} but has the advantage of being simpler and thus more adapted for the challenging environment offered by P4~\cite{chen2020implementing}. Also, it is not known if more complex and secure modes (CBC -- Cipher Block Chaining, CTR -- Counter Mode or GCM -- Galois Counter Mode) are efficiently implementable or feasible in P4, due to programming language limitations and hardware architecture constraints. 

\section{Implementing  RDMA and Encryption in P4} \label{sec:design} 

This section presents our practical implementation for secure, high-performance data transfers via RDMA, integrating AES-128 encryption on an Intel Tofino switch. 
We explain the AES encryption procedure and present the data plane implementation, discussing how to modify an existing AES 128-bit encryption solution to include RDMA headers and packet forwarding between hosts, limiting encryption to UDP and RDMA packets.

\subsection{Data Plane Implementation}

The data plane implementation builds upon an open-source solution\footnote{\url{https://github.com/Princeton-Cabernet/p4-projects/tree/master/AES-tofino}} for AES-128 encryption in P4~\cite{chen2020implementing}. This implementation was extended to support RDMA by adding parsing and forwarding capabilities for RoCEv2 packets. In detail, the switch was modified to handle RDMA-specific headers, enable packet forwarding between hosts, and selectively apply encryption to UDP/RDMA packets.

\subsubsection{Packet Types and Forwarding}
During the evaluation two types of packets were used: standard UDP packets and RDMA packets. UDP packets include Ethernet (14 bytes), IP (20 bytes), and UDP (8 bytes) headers. RDMA (RoCEv2) packets have the same base structure, with an additional 16 bytes comprising the BTH and ICRC headers. While this extra header information adds slight parsing overhead, it does not affect forwarding or encryption logic beyond initial header identification.
Packet forwarding in P4-enabled devices relies on programmable logic that determines packet handling based on header fields and metadata. In our implementation, we forwarded packets based on the recirculation decision. If encryption was complete, the packets were sent to the output port via the defined action \texttt{set\_egress\_port} that dynamically assigns the egress port based on parsing results, enabling flexible routing logic within the switch. Decisions were taken on a packet by packet basis through the use of match action tables that match header fields to actions; invalid packets were dropped.

The forwarding logic is integrated into the ingress pipeline via the ingress control block. This ensures that all packets are subjected to forwarding decisions immediately after parsing, and before they are forwarded or further processed within the switch.

\subsubsection{AES Encryption}


The scrambled lookup table technique, as used in the existing AES implementation that we leverage~\cite{chen2020implementing}, enables the deployment of AES on programmable switches, which do not support native cryptographic operations. This method exploits the large table memory of Reconfigurable Match Action Table based switches to replace complex arithmetic operations with table lookups, allowing the AES transformations to be mapped to memory-access patterns rather than sequential logic.
AES is a symmetric block cipher that encrypts 128-bit data blocks through multiple rounds of transformation. Each round comprises the following steps:
\begin{enumerate}[leftmargin=5.5mm]
    \item \textbf{AddRoundKey:} Each byte of the data block is XORed with the corresponding byte of the round key.
    \item \textbf{SubBytes:} Each byte is substituted using a nonlinear S-box lookup.
    \item \textbf{ShiftRows:} Rows of the 4x4 byte matrix are cyclically shifted by varying offsets.
    \item \textbf{MixColumns:} Each column is transformed using a fixed polynomial over a finite field to provide diffusion.
\end{enumerate}

These transformations are iterated over 10 rounds (for AES-128), with the final round omitting the MixColumns step. In the switch pipeline, most of these transformations are precomputed and stored as lookup tables, enabling the encryption to be performed entirely through memory accesses and basic logic.
%
%
The \texttt{AddRoundKey} and \texttt{SubBytes} steps were combined into a single lookup operation. This reduced the dependency chain and minimized the number of pipeline stages required.
Encryption keys were uploaded to the switch via a specific control-plane Python script. This script precomputes the required lookup tables for each AES round and uses the Barefoot runtime API to populate the switch’s match-action tables with these values.
To complete the full 10 rounds of AES-128, packets were recirculated through the switch pipeline multiple times. Each pipeline pass handled one encryption round, resulting in a total of 10 passes per packet. On the Intel Tofino switch, two internal ports were leveraged to support this recirculation mechanism efficiently.
Once encryption was complete, the packets were forwarded according to predefined rules. This ensured that encrypted traffic followed the appropriate egress path based on its destination and protocol.
The use of scrambled lookup tables enabled high throughput by reducing pipeline stage consumption. The memory-efficient design
allowed AES-128 encryption to be performed at approximately 11~Gbps on the Tofino switch.

\subsubsection{Variable Packet Sizes}

The original encryption logic was designed to handle a single 16-byte block per packet. This constraint significantly limited throughput and placed a high burden on end hosts to generate numerous small packets. Additionally, the majority of each packet consisted of protocol headers, leaving minimal room for useful payload data.
To address this, we extended the implementation to support the encryption of multiple 16-byte blocks per packet. The enhanced pipeline recirculates each packet for $10 \times N$ rounds, where $N$ is the number of 16-byte blocks in the payload. This required changes to the packet parser to support variable payload sizes and additional control logic to track encryption progress and determine when recirculation should terminate.

\section{Evaluation} \label{sec:evaluation}

This section outlines metrics and setups used to evaluate the trade-off between performance and security across diverse configurations.


\subsection{Evaluation Metrics}\label{metrics}

Throughput, packet loss, and latency are examined for different configurations. The objective is to explore the trade-off between performance and security, by measuring the decrease in throughput due to strengthened security. Understanding the impact of encryption on packet loss helps in assessing the feasibility and reliability of encrypted RDMA communication in practical deployment scenarios. 
To comprehensively assess the performance impact of RDMA with encryption, a multi-dimensional benchmarking strategy is followed. Benchmark evaluation is conducted using standardized tools such as Wireshark, Perftest, Qperf, Scapy, DPDK-Pktgen, TestPMD, and scripts to ensure repeatability and accuracy of results. The chosen benchmarks enable investigation of the following performance metrics:
Throughput quantifies the volume of data successfully transmitted per unit time and is calculated as:
\begin{equation}
\text{\textbf{Throughput} (Mb/s)} = \frac{\text{Total payload data transferred (Mb)}}{\text{Total time (s)}}
\end{equation}

We also measure the \textit{Mean Maximum Sustainable Throughput}, i.e. averaging multiple evaluation instances conducted for different payload size. 
The ``maximum sustainable throughput'' refers to the highest level of throughput that the system can maintain while keeping packet loss at a low level. Specifically, a cap was placed on the packet loss such that the value would be below 0.001\%, as was used in the 2019 Mellanox reference report from AMD. 

Packet loss reflects the reliability of the network, and is set as:
\begin{equation}
\text{\textbf{Packet Loss} (\%)} = \left(1 - \frac{\text{Received (RX) packets}}{\text{Transmitted (TX) packets}}\right) \times 100
\end{equation}

In scenarios where RDMA is deployed, packet loss can disrupt data transmission and compromise system performance.

\subsection{Evaluation Setups}
\begin{figure}[t]
  \setlength{\abovecaptionskip}{1pt}
  \setlength{\belowcaptionskip}{0pt}
  \centering
  \includegraphics[trim={1200 700 3270 1000 cm},clip,
                   width=0.65\linewidth]{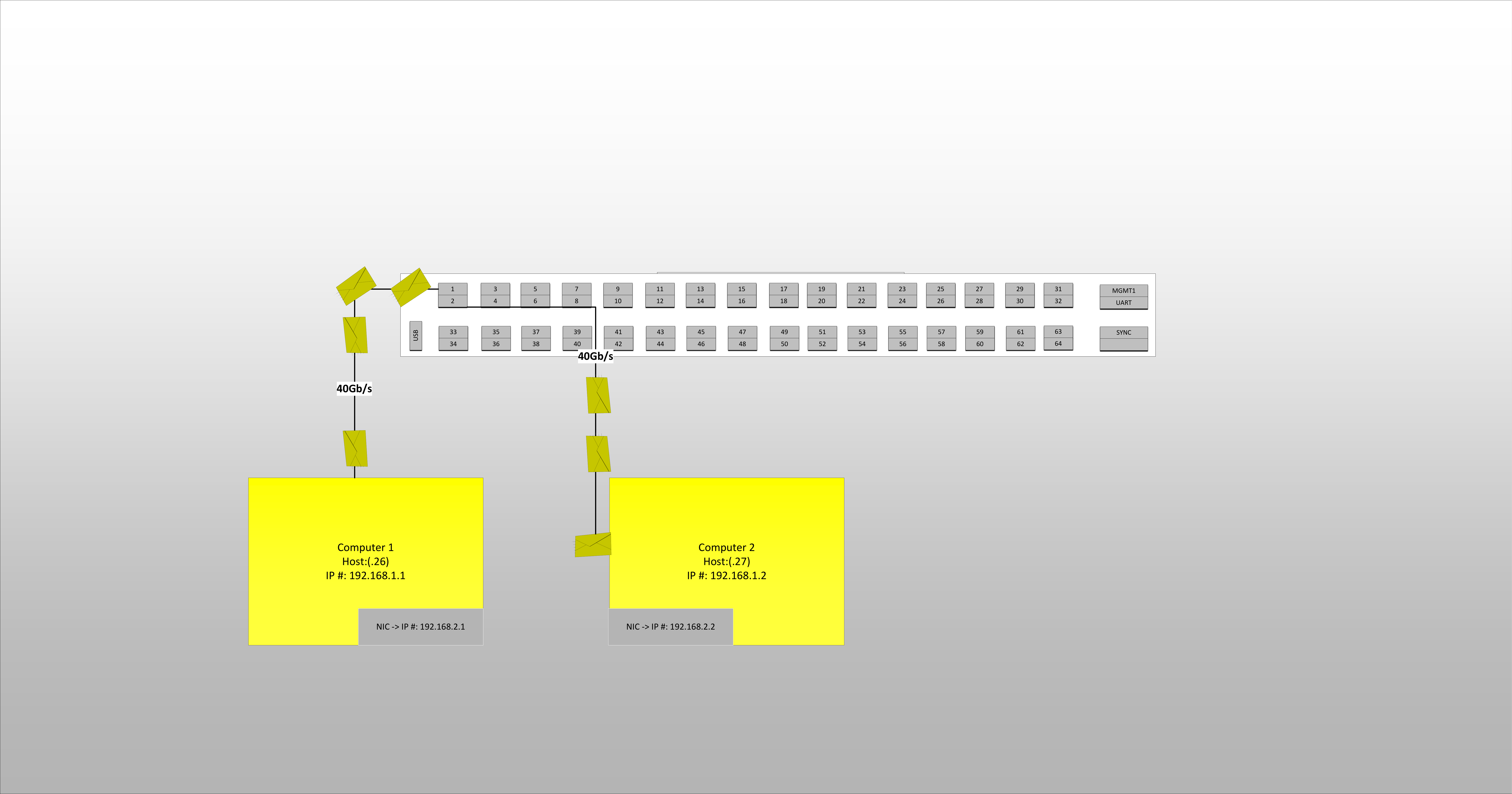}
  \caption{Lab Setup Topology}
  \label{fig:Topology}
\end{figure}

 Our hardware topology comprises a system consisting of two servers, Computer 1 (\textbf{C1}) and Computer 2 (\textbf{C2}) with 40 Gb/sec RDMA capable Mellanox ConnectX-3 nics. The servers are interconnected via a APS BF6064X-T Tofino switch, as illustrated in Figure \ref{fig:Topology}.
We decided to use two baseline setups and three experimental setups to evaluate the system. For our measurements, the baseline setups were constructed in such a way as to produce data that would demonstrate the behavior of the system without our proposed improvements. Similarly, the experimental setups were constructed with the intention of producing comparable data that would be able to showcase the difference in our chosen metrics compared to the baseline setups for the proposed system. 

\paragraph{Baseline setups}
The baseline configurations provide a foundation for assessing system performance with and without encryption, capturing metrics such as throughput and packet loss. These benchmarks enable direct comparisons to experimental setups, clarifying the impact of data plane encryption on overall performance.
\begin{itemize}
    \item \textbf{Baseline setup 1:} Generate unencrypted traffic through the switch and measure metrics (with and without RDMA).
    \item \textbf{Baseline setup 2:} Encrypt on Computer 1 and send traffic through the switch (without RDMA).
\end{itemize}

\paragraph{Experimental setups}
The experimental configurations evaluate the integration of encryption directly within the data plane. By embedding encryption within the switch, these setups help evaluate the switch's ability to handle high-throughput encrypted traffic while maintaining RDMA performance.
\begin{itemize}
    \item \textbf{Experimental setup 1:}
    Send DPDK traffic to switch and encrypt in the switch's dataplane.
    \item \textbf{Experimental setup 2:} 
    Generate traffic with a Python script and encrypt in the switch's dataplane.
     \item \textbf{Experimental setup 3:}
     Send \texttt{Ib\_send\_bw} and Qperf traffic to switch and encrypt in the switch's dataplane and compare.
\end{itemize}

\subsection{Experimental Pipeline}

\paragraph{CPU Implementation of Encryption/Decryption}

To enable unified evaluation across setups, AES-128 encryption and decryption were implemented in Python. C1 encrypts packets; C2 decrypts them.
On C1, a custom packet is constructed and encrypted using AES-128 in ECB mode via the \texttt{cryptography} library~\cite{cryptography}, with a fixed 128-bit key. The encrypted payload is then sent to C2 over UDP, then C2 receives the packet and decrypts it using the same key/algorithm to produce the output data. 

\paragraph{Packet Generation}

High-throughput packet transmission is demanding, especially for small packet sizes. As seen in the evaluation, Qperf alone could not demonstrate the full AES throughput on the Tofino switch due to inefficient small-packet generation. To address this, we used DPDK, which bypasses the kernel for faster user-space packet handling. This required a DPDK-compatible NIC and drivers.

\subsection{Results} \label{results}

We present here the results that were obtained in the evaluation. 

\subsubsection{Pilot Evaluation}
Preliminary measurements were conducted in a pilot study to establish typical performance figures in an unloaded scenario, the measurements were conducted between C1 and C2 through the switch. Throughput values of 21.6 Gb/sec Qperf TCP traffic and 37.2 Gb/sec for RDMA Write RC were achieved, when using a 40 Gb/sec NIC. The latency values were 8.95 $\mu$s for TCP and 0.74 $\mu$s for RDMA, demonstrating the clear advantages of using RDMA.

\subsubsection{Baseline Throughput and Message Rate}
\begin{figure}[b]
  \setlength{\abovecaptionskip}{1pt}
  \setlength{\belowcaptionskip}{0pt}
  \centering
  \includegraphics[width=0.95\linewidth]{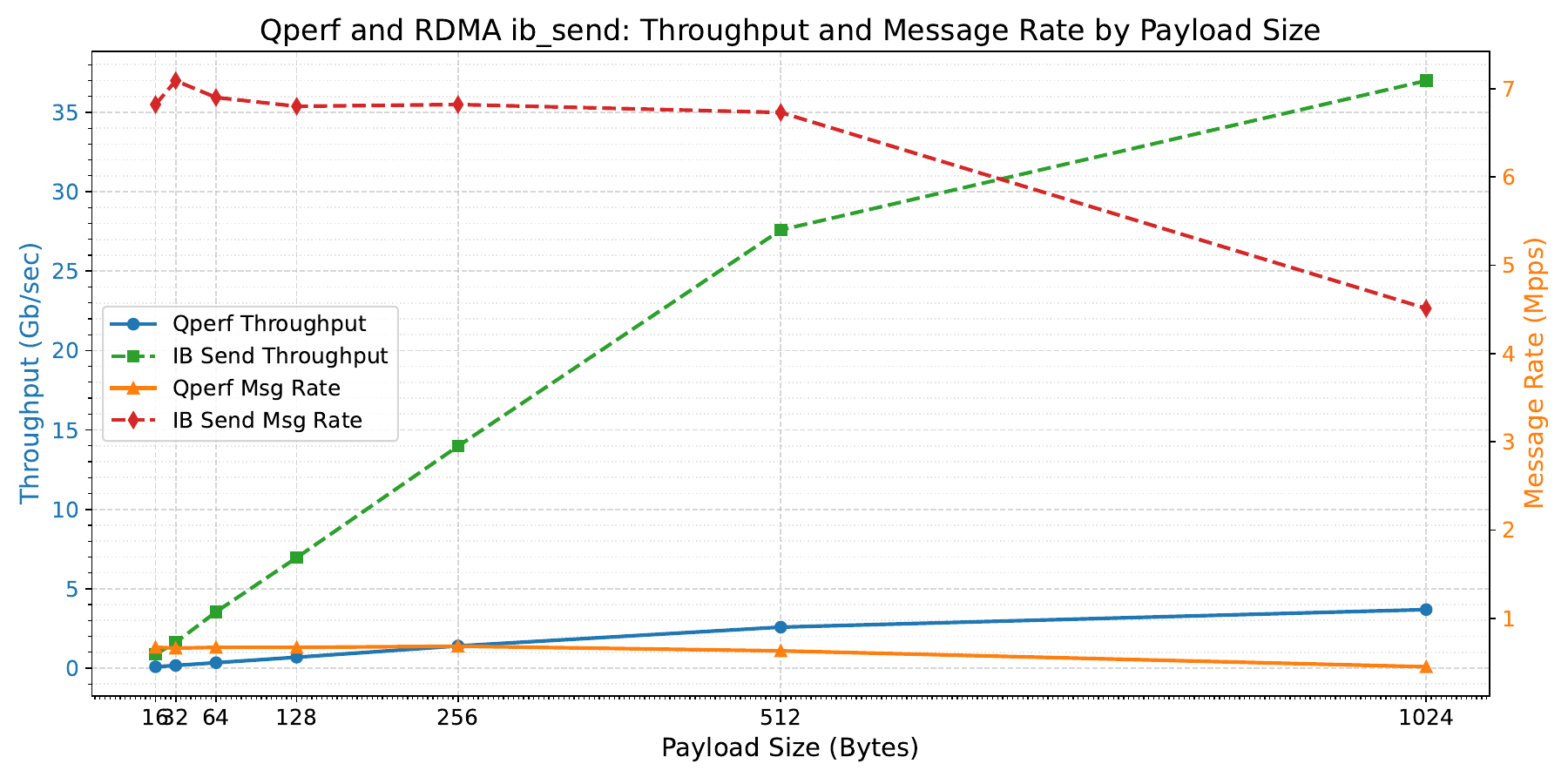}
  \caption{Baseline 1: throughput and message-rate trends versus payload size in RDMA Send and Qperf tests.}
  \label{fig:qperfvsibsendplot}
\end{figure}

Our baseline measurements are shown in Figure \ref{fig:qperfvsibsendplot} for RDMA Send UD packets and for Qperf UDP traffic, which shows that throughput increases with payload size, while message rate decreases slightly. The figure also highlights the performance advantage of RDMA.

\begin{figure}[t]
  \setlength{\abovecaptionskip}{1pt}
  \setlength{\belowcaptionskip}{0pt}
  \centering
  \includegraphics[width=0.90\linewidth]{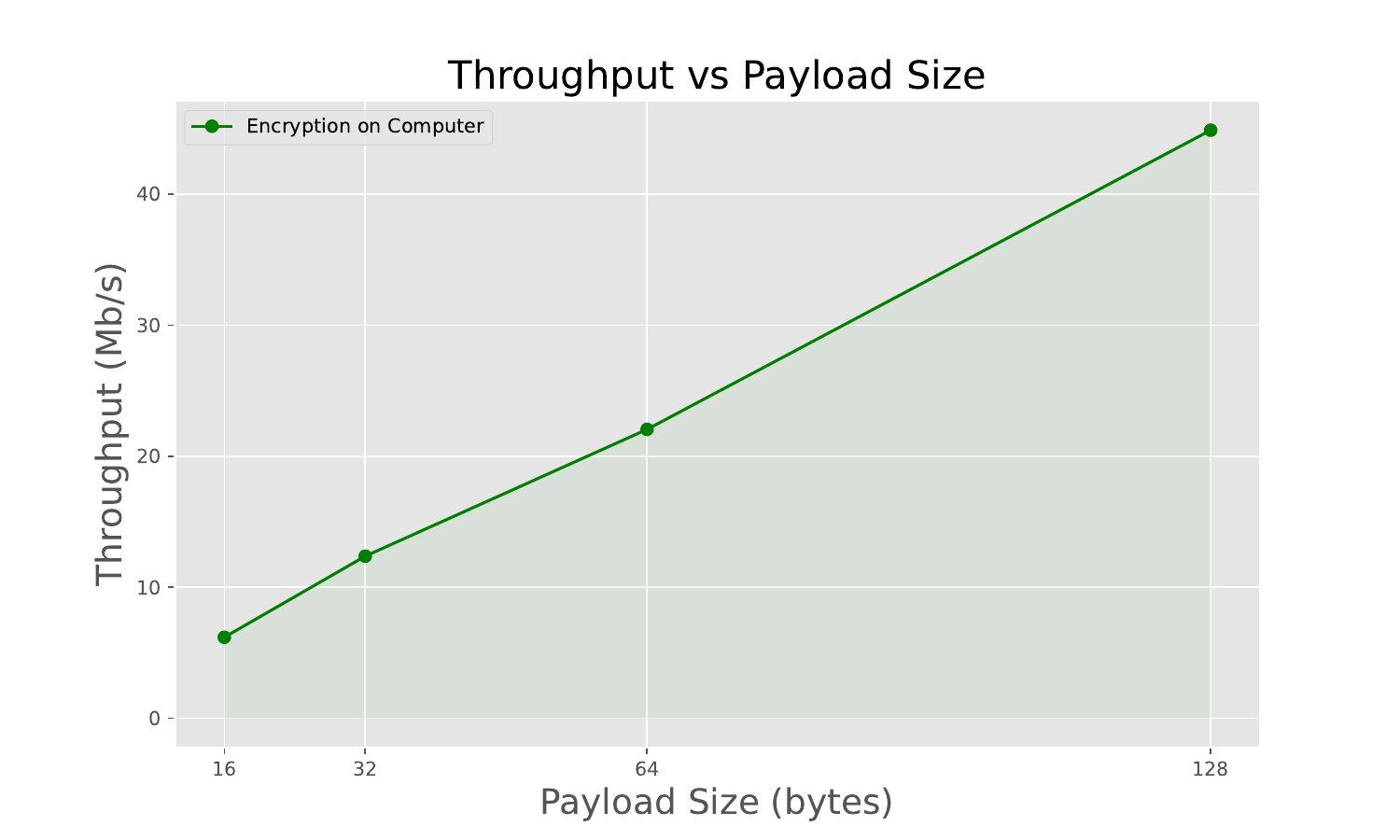}
  \caption{Baseline 2: throughput vs.\ payload size for encrypting on C1 and sending traffic through the switch (w/o RDMA).}
  \label{fig:Throughput_vs_Payload_Size_Baseline2}
\end{figure}

Figure \ref{fig:Throughput_vs_Payload_Size_Baseline2} shows that CPU-based encryption throughput increases with payload size but remains limited overall.

\subsubsection{DPDK-Pktgen Performance Evaluation}

\begin{figure}[b]
  \setlength{\abovecaptionskip}{1pt}
  \setlength{\belowcaptionskip}{0pt}
  \centering
  \includegraphics[width=0.95\linewidth]{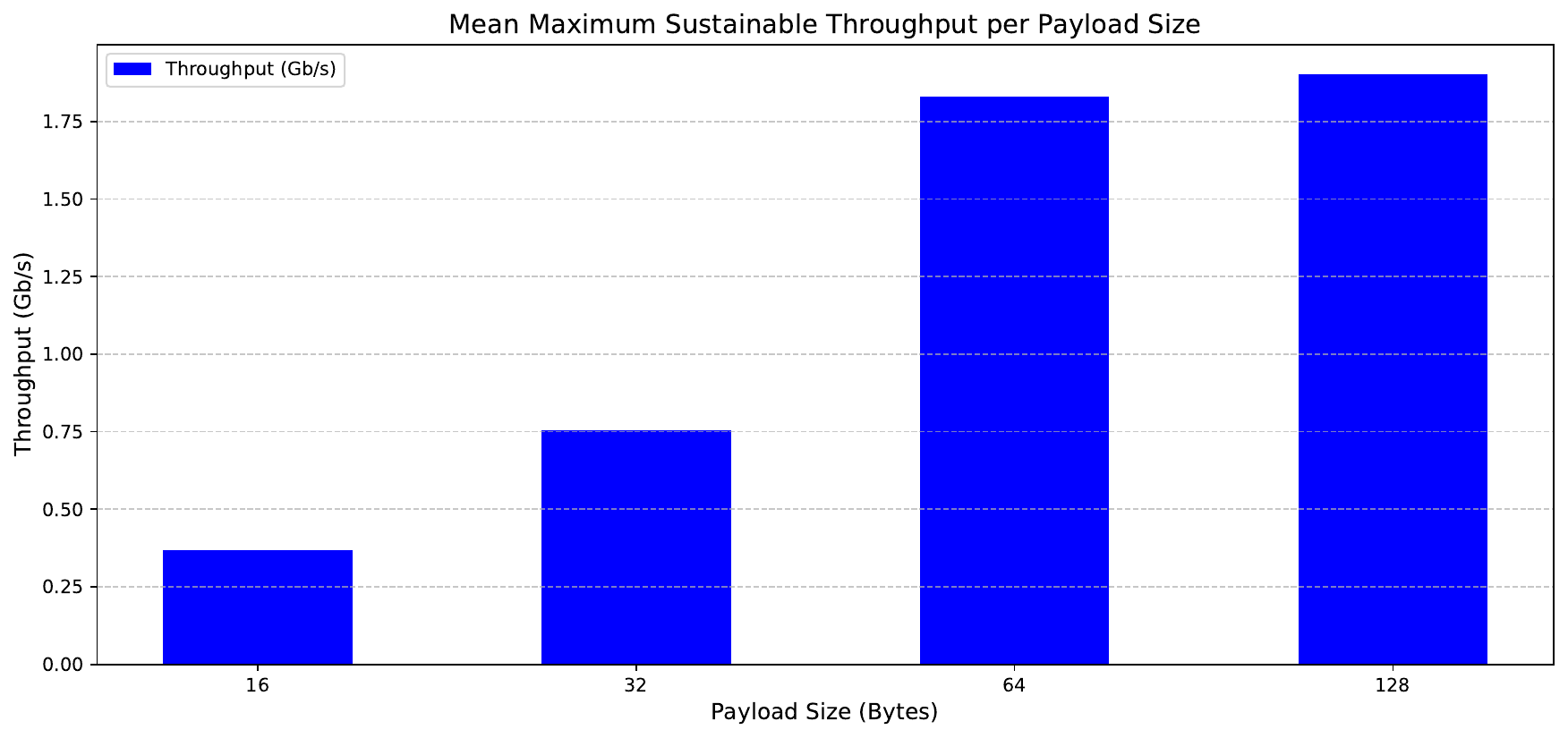}
  \caption{DPDK‐Pktgen performance metrics showing average throughput per payload size.}
  \label{fig:meanMaximumSustainableThroughput}
\end{figure}

Using DPDK-based tools for packet generation, we were able to drive high traffic rates even with smaller frames. The first evaluation of tests, therefore, examined how payload size affects our key metric—mean maximum sustainable throughput (see \S~\ref{metrics}). Figure \ref{fig:meanMaximumSustainableThroughput} shows a clear upward trend: as payloads grow, sustainable throughput rises. The reason is due to the larger packets amortizing protocol overhead across more user data and reducing the number of packets the system must process, easing both CPU and I/O burdens. Practically, this means that choosing larger payload sizes can improve bandwidth utilization and overall data-transfer performance.

\newcommand{\throughput}{\frac{\text{Total data transferred (bits)}}{\text{Total time (seconds)}}}

When transmitting 256-byte payloads, packet loss rose sharply even at low load, reaching 18\% at 0.03 Gbps, 38\% at 0.65 Gbps, and 97\% at 17.5 Gbps. This indicates that payloads above 128 bytes cannot be encrypted with acceptable loss in the current implementation. In addition, 384-byte payloads could not be compiled, suggesting that the maximum supported payload size had been reached.

\subsubsection{Performance Evaluation Vs CPU-based Encryption}
\begin{figure}[t]
  \setlength{\abovecaptionskip}{1pt}
  \setlength{\belowcaptionskip}{0pt}
  \centering
  \includegraphics[width=0.92\linewidth]{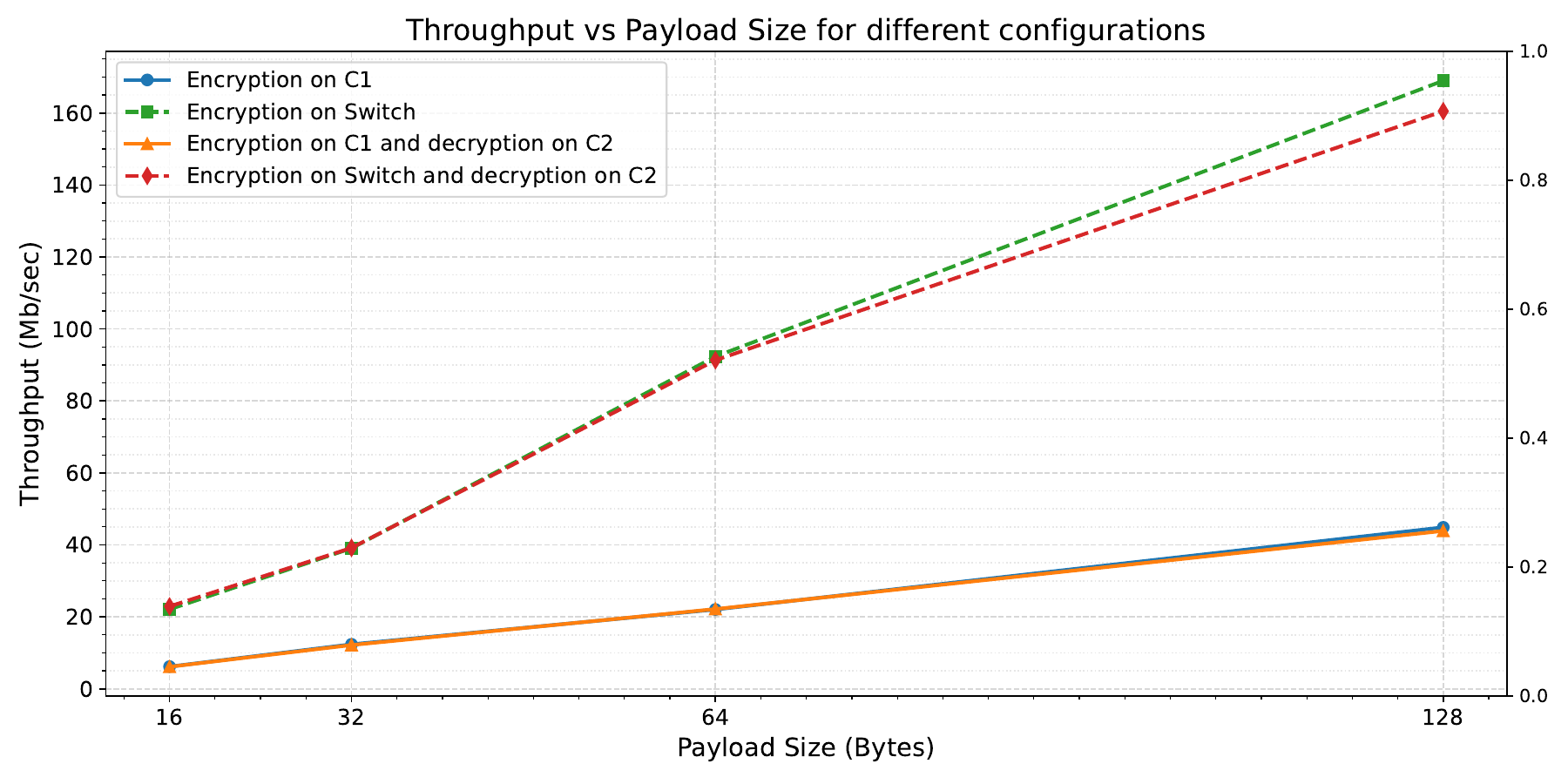}
  \caption{Throughput comparison for encryption on switch vs.\ computer, with/without decryption on C2.}
  \label{fig:throughputconfigurations}
\end{figure}

Integrating encryption within the switch provides substantial performance advantages compared to the computer-based encryption that was implemented with a Python script on C1. 
As seen in Figure \ref{fig:throughputconfigurations}, encryption on the switch achieves higher throughput across payload sizes and configurations. This offloading reduces CPU usage, as the CPU would otherwise be involved in both encryption and packet generation. The switch processes encryption directly, bypassing the CPU, and lowering overhead for faster data transfer rates. For example, with a payload size of 128 bytes, the switch achieves a throughput of approximately 168.98 Mb/s, vastly outperforming the computer's 44.88 Mb/s.
Both methods show increased packet loss as payload size grows, but switch-based encryption incurs slightly higher loss at larger payloads (e.g., 11.2\% vs. 9.32\% at 128 bytes). Despite this, its significantly higher throughput indicates more efficient handling of larger volumes of encrypted data. The rise in packet loss at higher payloads and throughput in both cases is explained by the decryption process in C2 being the bottleneck.

\begin{figure}[t]
  \setlength{\abovecaptionskip}{1pt}
  \setlength{\belowcaptionskip}{0pt}
  \centering
  \includegraphics[width=0.92\linewidth]{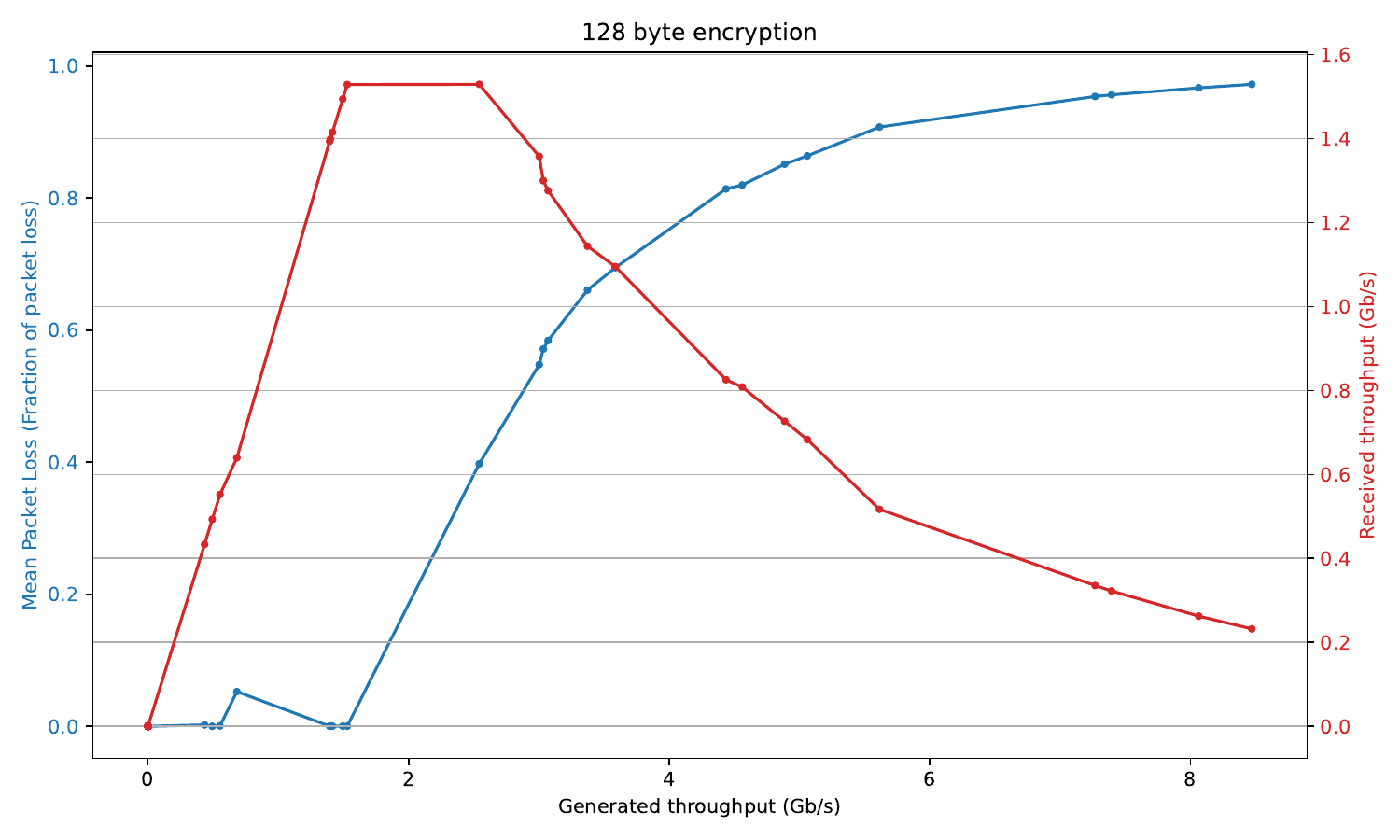}
  \caption{Received throughput on C2 as a function of generated packets on C1.}
  \label{fig:RxVsTx}
\end{figure}

Figure \ref{fig:RxVsTx} shows the throughput received in C2 as a function of the generated throughput from C1. Ideally, the two should match, while deviations indicate congestion or hardware bottlenecks. In this case, the switch became the bottleneck at higher rates, when encrypting.

\subsubsection{Summary of Results and Discussion}

Our AES implementation on the Tofino switch achieved up to 1.9 Gbps for 128-byte payloads—competitive compared to host-based encryption but limited by hardware constraints. Parsing each 16-byte block separately likely bottlenecks performance, and added control logic restricts the number of rounds per recirculation. Optimizations, such as using the egress pipeline or dedicating physical ports to recirculation, may improve throughput.

\section{Conclusion and Outlook} \label{sec:conclusion}
This work demonstrates that AES encryption for RDMA traffic can be implemented entirely in the data plane. By offloading cryptographic processing to the switch, our prototype achieved promising throughput for small to mid-sized payloads.

Our results show that offloading AES-128 encryption to the data plane allows for a secure data transfer while still maintaining a tolerable RDMA performance. In particular, we achieved throughput values of 0.37 Gbps for 16-byte packets, 0.76 Gbps for 32-byte packets, 1.83 Gbps for 64-byte packets, and 1.90 Gbps for 128-byte packets when encrypting in the switch. 
Packet loss remained negligible below these ceilings. Larger payloads tolerated higher injection rates, whereas bursts of small packets exposed the full cost of encryption. Our evaluation showed robust performance for payloads up to 128 bytes. However, packet loss increased at 256 bytes, and 384-byte payloads could not be encrypted due to architectural constraint of the switch.
 
 Tofino-based encryption remains limited in terms of security. It lacks secure modes (e.g., GCM, CTR), forward secrecy, and protected host-switch links. 
 Key areas for future work are implementing AES-CTR, enabling in-switch decryption, and implementing robust key exchange protocols to securely establish the initial encrypted session, while ensuring forward secrecy of transmitted data.
 
\bibliographystyle{ACM-Reference-Format}
\bibliography{references}

\end{document}